\documentclass[11pt]{article}
\usepackage{amssymb,amsmath,amsfonts}
\usepackage{graphicx}
\usepackage{graphics}
\usepackage{eepic,epsfig}

\begin{document}

\title{Induced cosmological constant in braneworlds \\ with warped internal
spaces}
\author{Aram A. Saharian \thanks{%
E-mail: saharyan@server.physdep.r.am } \\
\\
\textit{Department of Physics, Yerevan State University, 1 Alex Manoogian
Str.,}\\
\textit{375025 Yerevan, Armenia} \\
\textit{and}\\
\textit{Departamento de F\'{\i}sica-CCEN, Universidade Federal da Para\'{\i}%
ba,} \\
\textit{\ 58.059-970, Caixa Postal 5.008, Jo\~{a}o Pessoa, PB, Brazil}}
\maketitle

\begin{abstract}
We investigate the vacuum energy density induced by quantum
fluctuations of a bulk scalar field with general curvature
coupling parameter on two codimension one parallel branes in a
$(D+1)$-dimensional background spacetime
${\mathrm{AdS}}_{D1+1}\times \Sigma $ with a warped internal space
$\Sigma $. It is assumed that on the branes the field obeys Robin
boundary conditions. Using the generalized zeta function technique
in combination with contour integral representations, the surface
energies on the branes are presented in the form of the sums of
single brane and second brane induced parts. For the geometry of a
single brane both regions, on the left (L-region) and on the right
(R-region), of the brane are considered. The surface densities for
separate L- and R-regions contain pole and finite contributions.
For an infinitely thin brane taking these regions together, in odd
spatial dimensions the pole parts cancel and the total surface
energy is finite. The parts in the surface densities generated by
the presence of the second brane are finite for all nonzero values
of the interbrane separation. The contribution of the Kaluza-Klein
modes along $\Sigma $ is investigated in various limiting cases.
It is shown that for large distances between the branes the
induced surface densities give rise to an exponentially suppressed
cosmological constant on the brane. In the higher dimensional
generalization of the Randall-Sundrum braneworld model, for the
interbrane distances solving the hierarchy problem, the
cosmological constant generated on the visible brane is of the
right order of magnitude with the value suggested by the
cosmological observations.
\end{abstract}

\bigskip


\section{Introduction}

\label{sec:introd}

Motivated by the problems of the radion stabilization and the
generation of cosmological constant, the role of quantum effects
in braneworlds has attracted great deal of attention
\cite{Fabi00}-\cite{Saha06d}. A class of higher dimensional models
with the topology $\mathrm{AdS}_{D_{1}+1}\times \Sigma $, where
$\Sigma $ is a one-parameter compact manifold, and with two branes
of codimension one located at the orbifold fixed points, is
considered in Refs. \cite{Flac03b,Flac03}. In both cases of the
warped and unwarped internal manifold, the quantum effective
potential induced by bulk scalar fields is evaluated and it has
been shown that this potential can stabilize the hierarchy between
the Planck and electroweak scales without fine tuning. In addition
to the effective potential, the investigation of local physical
characteristics in these models is of considerable interest. Local
quantities contain more information on the vacuum fluctuations
than the global ones and play an important role in modelling a
self-consistent
dynamics involving the gravitational field. In the previous papers \cite%
{Saha06a,Saha06b} we have studied the Wightman function, the vacuum
expectation values of the field square and the energy-momentum tensor for a
scalar field with an arbitrary curvature coupling parameter obeying Robin
boundary conditions on two codimension one parallel branes embedded in the
background spacetime $\mathrm{AdS}_{D_{1}+1}\times \Sigma $ with a warped
internal space $\Sigma $. For an arbitrary internal space $\Sigma $, the
application of the generalized Abel-Plana formula \cite{SahaRev} allowed us
to extract form the vacuum expectation values the part due to the bulk
without branes and to present the brane induced parts in terms of
exponentially convergent integrals for the points away from the branes.

The braneworld corresponds to a manifold with boundaries and the physical
quantities, in general, receive both volume and surface contributions. In
particular, the contributions located on the visible brane are of special
interest as they are direct observables in the theory. In Ref.~\cite%
{Saha04surf} the vacuum expectation value of the surface
energy-momentum tensor is evaluated for a massive scalar field
subject to Robin boundary conditions on two parallel branes in
$(D+1)$-dimensional AdS bulk. It has been shown that for large
distances between the branes the induced surface densities give
rise to an exponentially suppressed cosmological constant on the
brane. In the Randall-Sundrum braneworld model, for the interbrane
distances solving the hierarchy problem between the gravitational
and electroweak mass scales, the cosmological constant generated
on the visible brane is of the right order of magnitude with the
value suggested by the cosmological observations. In the present
talk based on \cite{Saha06d}, we describe similar issues within
the framework of higher dimensional braneworld models with warped
internal spaces.

\section{Surface energy-momentum tensor and the zeta function}

\label{sec:zetafunc}

Consider a scalar field $\varphi (x)$ with curvature coupling
parameter $\zeta $ satisfying the equation of motion
\begin{equation}
\left( \nabla ^{M}\nabla _{M}+m^{2}+\zeta R\right) \varphi (x)=0,
\label{fieldeq}
\end{equation}%
where $\nabla _{M}$ is the covariant derivative operator and $R$ is the
scalar curvature for a $(D+1)$-dimensional background spacetime. We will
assume that the bulk has the topology $\mathrm{AdS}%
_{D_{1}+1}\times \Sigma $ and is described by the line element
\begin{equation}
ds^{2}=g_{MN}dx^{M}dx^{N}=e^{-2k_{D}y}\left( \eta _{\mu \sigma }dx^{\mu
}dx^{\sigma }-\gamma _{ik}dX^{i}dX^{k}\right) -dy^{2},  \label{metric}
\end{equation}%
where $\eta _{\mu \sigma }$ is the metric tensor for $D_{1}$-dimensional
Minkowski spacetime, $k_{D}$ is the inverse AdS radius, and the coordinates $%
X^{i}$, $i=1,\ldots ,D_{2}$, cover the internal manifold $\Sigma $, $%
D=D_{1}+D_{2}$. The Ricci scalar corresponding \ to line element (\ref%
{metric}) is given by formula $R=-D(D+1)k_{D}^{2}-e^{2k_{D}y}R_{(\gamma )}$,
with $R_{(\gamma )}$ being the scalar curvature for the metric tensor $%
\gamma _{ik}$. In the discussion below, in addition to the radial coordinate
$y$ we will also use the coordinate $z=e^{k_{D}y}/k_{D}$, in terms of which
the line element is written in the form conformally related to the metric in
the direct product spacetime $R^{(D_{1},1)}\times \Sigma $ by the conformal
factor $(k_{D}z)^{-2}$.

We are interested in one-loop vacuum effects induced by quantum fluctuations
of the bulk field $\varphi (x)$ on two parallel branes of codimension one,
located at $y=a$ and $y=b$, $a<b$. We assume that on the branes the field
obeys Robin boundary conditions
\begin{equation}
\left( \tilde{A}_{y}+\tilde{B}_{y}\partial _{y}\right) \varphi (x)=0,\quad
y=a,b,  \label{boundcond}
\end{equation}%
with constant coefficients $\tilde{A}_{y}$, $\tilde{B}_{y}$. In a
higher dimensional generalization of the Randall-Sundrum
braneworld model the
coordinate $y$ is compactified on an orbifold $S^{1}/Z_{2}$, of length $l$, $%
-l\leqslant y\leqslant l$, and the orbifold fixed points $y=0$ and $y=l$ are
the locations of two branes. The corresponding line-element has the form (%
\ref{metric}) with the warp factor $e^{-2k_{D}|y|}$. In these models the
region between the branes is employed only. For an untwisted bulk scalar
with brane mass terms $c_{a}$ and $c_{b}$, the corresponding ratio of the
coefficients in the boundary condition (\ref{boundcond}) is determined by
the expression (see, e.g., Refs. \cite{Gher00,Flac01b,Saha05} for the case
of the bulk $\mathrm{AdS}_{D+1}$ and Refs. \cite{Flac03b,Saha06a} for the
geometry under consideration)
\begin{equation}
\frac{\tilde{A}_{j}}{\tilde{B}_{j}}=-\frac{n^{(j)}c_{j}+4D\zeta k_{D}}{2}%
,\;n^{(a)}=1,\;n^{(b)}=-1.  \label{AjBjRS}
\end{equation}%
In the supersymmetric version of the model \cite{Gher00} one has $%
c_{b}=-c_{a}$ and the boundary conditions are the same for both branes. For
a twisted scalar, Dirichlet boundary conditions are obtained on both branes.

For the geometry of two parallel branes in $\mathrm{AdS}_{D_{1}+1}\times
\Sigma $ with boundary conditions (\ref{boundcond}), the Wightman function
and the vacuum expectation values (VEVs) of the field square and the bulk
energy-momentum tensor are investigated in Refs. \cite{Saha06a,Saha06b}. On
manifolds with boundaries the energy-momentum tensor in addition to the bulk
part contains a contribution located on the boundary. For an arbitrary
smooth boundary $\partial M$ with the inward-pointing unit normal vector $%
n^{L}$, the surface part of the energy-momentum tensor for a
scalar field \cite{Saha03emt} is given by the formula
$T_{MN}^{\mathrm{(s)}}=\delta (x;\partial M)\tau _{MN}$, where the
"one-sided" delta-function $\delta (x;\partial M)$ locates this
tensor on $\partial M$ and
\begin{equation}
\tau _{MN}=\zeta \varphi ^{2}K_{MN}-(2\zeta -1/2)h_{MN}\varphi n^{L}\nabla
_{L}\varphi .  \label{tausurf}
\end{equation}%
In Eq. (\ref{tausurf}), $h_{MN}=g_{MN}+n_{M}n_{N}$ is the induced metric on
the boundary and $K_{MN}=h_{M}^{L}h_{N}^{P}\nabla _{L}n_{P}$ is the
corresponding extrinsic curvature tensor.

By expanding the field operator over a complete set of eigenfunctions $%
\{\varphi _{\alpha }(x),\varphi _{\alpha }^{\ast }(x)\}$\ obeying the
boundary conditions and using the standard commutation rules, for the VEV of
the operator $\tau _{MN}$ one finds $\langle 0|\tau _{MN}|0\rangle
=\sum_{\alpha }\tau _{MN}\{\varphi _{\alpha }(x),\varphi _{\alpha }^{\ast
}(x)\}$, where the bilinear form $\tau _{MN}\{\varphi ,\psi \}$ is
determined by the classical energy-momentum tensor (\ref{tausurf}). Below in
this section, we will consider the region between the branes. The
corresponding quantities for the regions $y\leqslant a$ and $y\geqslant b$
are obtained as limiting cases. As we have mentioned before, in the
orbifolded version of the model, the only bulk is the one between the
branes. In this region the inward-pointing normal to the brane at $y=j$, $%
j=a,b$, and the corresponding extrinsic curvature tensor are given by the
relations $n^{(j)M}=n^{(j)}\delta _{D}^{M}$ and $%
K_{MN}^{(j)}=-n^{(j)}k_{D}h_{MN}$, where $n^{(j)}$ is defined in formula (%
\ref{AjBjRS}). By using these relations and the boundary conditions, the VEV
of the surface energy-momentum tensor on the brane at $y=j$ is presented in
the form
\begin{equation}
\langle 0|\tau _{MN}^{(j)}|0\rangle =-h_{MN}n^{(j)}k_{D}C_{j}\langle
0|\varphi ^{2}|0\rangle _{y=j},\;C_{j}\equiv \zeta -(2\zeta -1/2)\tilde{A}%
_{j}/(k_{D}\tilde{B}_{j}).  \label{tauj}
\end{equation}%
From the point of view of physics on the brane, Eq. (\ref{tauj}) corresponds
to the gravitational source of the cosmological constant type with the
surface energy density $\varepsilon _{j}^{{\mathrm{(s)}}}=\langle 0|\tau
_{0}^{(j)0}|0\rangle $ (surface energy per unit physical volume on the brane
at $y=j$ or brane tension), stress $p_{j}^{{\mathrm{(s)}}}=-\langle 0|\tau
_{1}^{(j)1}|0\rangle $, and the equation of state $\varepsilon _{j}^{{%
\mathrm{(s)}}}=-p_{j}^{{\mathrm{(s)}}}$. It is remarkable that
this relation takes place for both subspaces on the brane. It can
be seen that this result is valid also for the general metric
$g_{\mu \sigma }$ instead of $\eta _{\mu \sigma }$ in line element
(\ref{metric}). For an untwisted bulk scalar in the higher
dimensional generalization of the Randall-Sundrum braneworld based
on the bulk ${\mathrm{AdS}}_{D_{1}+1}\times \Sigma $ the
coefficient in Eq. (\ref{tauj}) is given by the formula
\begin{equation}
C_{j}=\zeta (4D\zeta -D+1)+(\zeta -1/4)n^{(j)}c_{j}/k_{D},  \label{gorcRS}
\end{equation}%
In particular, the corresponding surface energy vanishes for minimally and
conformally coupled scalar fields with zero brane mass terms.

In order to evaluate the expectation value of the surface
energy-momentum tensor we need the corresponding eigenfunctions.
These functions can be taken in the decomposed form
\begin{equation}
\varphi _{\alpha }(x^{M})=\frac{f_{n}(y)e^{-i\eta _{\mu \nu }k^{\mu }x^{\nu
}}}{\sqrt{2\omega _{\beta ,n}(2\pi )^{D_{1}-1}}}\psi _{\beta }(X),\;k^{\mu
}=(\omega _{\beta ,n},\mathbf{k}),\;\omega _{\beta ,n}=\sqrt{%
k^{2}+m_{n}^{2}+\lambda _{\beta }^{2}},\;k=|\mathbf{k}|,  \label{eigfunc1}
\end{equation}%
where the mass spectrum $m_{n}$ is determined by the boundary conditions.
The modes $\psi _{\beta }(X)$ are eigenfunctions for the internal subspace:%
\begin{equation}
\left[ \Delta _{(\gamma )}+\zeta R_{(\gamma )}\right] \psi _{\beta
}(X)=-\lambda _{\beta }^{2}\psi _{\beta }(X),  \label{psibeteq}
\end{equation}%
with the eigenvalues $\lambda _{\beta }^{2}$ and the orthonormalization
condition $\int d^{D_{2}}X\,\sqrt{\gamma }\psi _{\beta }(X)\psi _{\beta
^{\prime }}^{\ast }(X)=\delta _{\beta \beta ^{\prime }}$. In Eq. (\ref%
{psibeteq}), $\Delta _{(\gamma )}$ is the Laplace-Beltrami operator for the
metric $\gamma _{ij}$. From the field equation (\ref{fieldeq}), one obtains
the equation for the function $f_{n}(y)$ with the solution
\begin{equation}
f_{n}(y)=C_{n}e^{Dk_{D}y/2}g_{\nu }^{(a)}(m_{n}z_{a},m_{n}z),\;g_{\nu
}^{(j)}(u,v)\equiv J_{\nu }(v)\bar{Y}_{\nu }^{(j)}(u)-Y_{\nu }(v)\bar{J}%
_{\nu }^{(j)}(u),\;j=a,b,  \label{fny}
\end{equation}%
where the normalization coefficient $C_{n}$ is determined below. Here $%
J_{\nu }(x)$, $Y_{\nu }(x)$ are the Bessel and Neumann functions of the
order
\begin{equation}
\nu =\sqrt{D^{2}/4-D(D+1)\zeta +m^{2}/k_{D}^{2}},  \label{nu}
\end{equation}%
and the $z$-coordinates of the branes are denoted by $z_{j}=e^{k_{D}j}/k_{D}$%
, $j=a,b$. In formula (\ref{fny}) and in what follows for a given function $%
F(x)$ we use the notation
\begin{equation}
\bar{F}^{(j)}(x)=A_{j}F(x)+B_{j}xF^{\prime }(x),\;A_{j}=\tilde{A}_{j}+\tilde{%
B}_{j}k_{D}D/2,\;B_{j}=\tilde{B}_{j}k_{D},\;j=a,b.  \label{notbar}
\end{equation}%
Note that for conformally and minimally coupled massless scalar fields one
has $\nu =1/2$ and $\nu =D/2$, respectively.

The function (\ref{fny}) satisfies the boundary condition on the brane at $%
y=a$. Imposing the boundary condition on the brane $y=b$ we find that the
eigenvalues $m_{n}$ are solutions to the equation
\begin{equation}
g_{\nu }^{(ab)}(m_{n}z_{a},m_{n}z_{b})\equiv \bar{J}_{\nu }^{(a)}(m_{n}z_{a})%
\bar{Y}_{\nu }^{(b)}(m_{n}z_{b})-\bar{Y}_{\nu }^{(a)}(m_{n}z_{a})\bar{J}%
_{\nu }^{(b)}(m_{n}z_{b})=0.  \label{cnu}
\end{equation}%
Equation (\ref{cnu}) determines the Kaluza-Klein (KK) spectrum along the
transverse dimension. We denote by $u=\gamma _{\nu ,n}$, $n=1,2,\ldots $,
the positive zeros of the function $g_{\nu }^{(ab)}(u,uz_{b}/z_{a})$,
arranged in the ascending order, $\gamma _{\nu ,n}<\gamma _{\nu ,n+1}$. The
eigenvalues for $m_{n}$ are related to these zeros as $m_{n}=\gamma _{\nu
,n}/z_{a}$. From the orthonormality condition of the radial functions, for
the coefficient $C_{n}$ in Eq. (\ref{fny}) one finds
\begin{equation}
C_{n}^{2}=\frac{\pi m_{n}}{k_{D}}\frac{\bar{Y}_{\nu }^{(b)}(m_{n}z_{b})/\bar{%
Y}_{\nu }^{(a)}(m_{n}z_{a})}{\frac{\partial }{\partial u}g_{\nu
}^{(ab)}(uz_{a},uz_{b})|_{u=m_{n}}}.  \label{cn}
\end{equation}%
Note that, as we consider the quantization in the region between the branes,
$a\leqslant y\leqslant b$, the modes defined by (\ref{fny}) are normalizable
for all real values of $\nu $ from Eq. (\ref{nu}).

Substituting the eigenfunctions (\ref{eigfunc1}) into the corresponding mode
sum and integrating over the angular part of the vector $\mathbf{k}$, for
the expectation value of the energy density on the brane at $y=j$ we obtain
\begin{equation}
\varepsilon _{j}^{\text{\textrm{(s)}}}=-2n^{(j)}C_{j}k_{D}^{D}z_{j}^{D}B_{j}%
\beta _{D_{1}-1}\int_{0}^{\infty }dk\,k^{D_{1}-2}\sum_{\beta
}\sum_{n=1}^{\infty }\frac{|\psi _{\beta }(X)|^{2}m_{n}g_{\nu
}^{(l)}(m_{n}z_{l},m_{n}z_{j})}{\omega _{\beta ,n}\left. \frac{\partial }{%
\partial u}g_{\nu }^{(ab)}(uz_{a},uz_{b})\right\vert _{u=m_{n}}},
\label{phi2j}
\end{equation}%
where $j,l=a,b$, and $l\neq j$, and%
\begin{equation}
\beta _{D_{1}}=\frac{1}{(4\pi )^{D_{1}/2}\Gamma \left( D_{1}/2\right) }.
\label{betaD}
\end{equation}%
To regularize the divergent expression on the right of this formula we
define the function
\begin{equation}
\Phi _{j}(s)=2z_{j}^{D}\frac{B_{j}\beta _{D_{1}-1}}{\mu ^{1+s}}\sum_{\beta
}|\psi _{\beta }(X)|^{2}\int_{0}^{\infty }dk\,k^{D_{1}-2}\sum_{n=1}^{\infty
}\omega _{\beta ,n}^{s}\frac{m_{n}g_{\nu }^{(l)}(m_{n}z_{l},m_{n}z_{j})}{%
\frac{\partial }{\partial u}g_{\nu }^{(ab)}(uz_{a},uz_{b})|_{u=m_{n}}},
\label{IAs}
\end{equation}%
where an arbitrary mass scale $\mu $\ is introduced to keep the dimension of
the expression. After the evaluation of the integral over $k$, this
expression can be presented in the form%
\begin{equation}
\Phi _{j}(s)=\frac{z_{j}^{D}B_{j}}{(4\pi )^{(D_{1}-1)/2}}\sum_{\beta }|\psi
_{\beta }(X)|^{2}\zeta _{j\beta }(s),  \label{Fjs2}
\end{equation}%
where the generalized partial zeta function
\begin{equation}
\zeta _{j\beta }(s)=\frac{\Gamma (-\alpha _{s})}{\Gamma (-s/2)\mu ^{s+1}}%
\sum_{n=1}^{\infty }(m_{n}^{2}+\lambda _{\beta }^{2})^{\alpha _{s}}\frac{%
m_{n}g_{\nu }^{(l)}(m_{n}z_{l},m_{n}z_{j})}{\frac{\partial }{\partial u}%
g_{\nu }^{(ab)}(uz_{a},uz_{b})|_{u=m_{n}}},\;\alpha _{s}=\frac{D_{1}+s-1}{2}.
\label{zetsx}
\end{equation}%
The computation of the VEV of the surface energy-momentum tensor requires
the analytic continuation of the function $\Phi _{j}(s)$ to the value $s=-1$%
. In order to obtain this analytic continuation we will follow the procedure
multiply used for the evaluation of the Casimir energy (see, for instance,
\cite{Bord01}). On the base of this procedure it can be seen that in the
strip $-D_{1}-1<{\mathrm{Re}}\,s<-D_{1}$ of the complex plane $s$ we have
the following integral representation
\begin{equation}
\zeta _{j\beta }(s)=\zeta _{j\beta }^{\mathrm{(J)}}(s)-\frac{\mu ^{-s-1}B_{j}%
}{\Gamma \left( -\frac{s}{2}\right) \Gamma (\alpha _{s}+1)}\int_{\lambda
_{\beta }}^{\infty }du\,u(u^{2}-\lambda _{\beta }^{2})^{\alpha _{s}}\Omega
_{j\nu }(uz_{a},uz_{b}),  \label{zet12}
\end{equation}%
where
\begin{equation}
\zeta _{j\beta }^{\mathrm{(J)}}(s)=-\frac{n^{(j)}\mu ^{-s-1}}{\Gamma
(-s/2)\Gamma (\alpha _{s}+1)}\int_{\lambda _{\beta }}^{\infty
}du\,u(u^{2}-\lambda _{\beta }^{2})^{\alpha _{s}}\frac{F_{\nu }(uz_{j})}{%
\bar{F}_{\nu }^{(j)}(uz_{j})}.  \label{FLs}
\end{equation}%
In (\ref{zet12}) we have defined
\begin{equation}
\Omega _{a\nu }(u,v)=\frac{\bar{K}_{\nu }^{(b)}(v)}{\bar{K}_{\nu
}^{(a)}(u)G_{\nu }^{(ab)}(u,v)},\;\Omega _{b\nu }(u,v)=\frac{\bar{I}_{\nu
}^{(a)}(u)}{\bar{I}_{\nu }^{(b)}(v)G_{\nu }^{(ab)}(u,v)},  \label{Omegab}
\end{equation}%
with the modified Bessel functions $I_{\nu }(u)$ and $K_{\nu }(u)$. In
formula (\ref{FLs}) we use the notation $F=K$ for $j=a$ and $F=I$ for $j=b$.
The contribution of the second term on the right of Eq. (\ref{zet12}) is
finite at $s=-1$ and vanishes in the limits $z_{a}\rightarrow 0$ or $%
z_{b}\rightarrow \infty $. The first term corresponds to the contribution of
a single brane at $z=z_{j}$ when the second brane is absent. The surface
energy density corresponding to this term is located on the surface $y=a+0$
for the brane at $y=a$ and on the surface $y=b-0$ for the brane at $y=b$. To
distinguish on which side of the brane is located the corresponding term we
use the superscript J=L for the left side and J=R for the right side. As we
consider the region between the branes, in formula (\ref{FLs}) J=R for $j=a$
and J=L for $j=b$. The further analytic continuation is needed for the
function $\zeta _{j\beta }^{\mathrm{(J)}}(s)$ only and this is done in the
next section. The integral representation for the partial zeta function
given by Eqs. (\ref{zet12}), (\ref{FLs}) is valid for the case of the
presence of the zero mode as well.

\section{Surface energy density for a single brane}

\label{sec:1brane}

In this section we consider the geometry of a single brane placed at $y=j$.
The orbifolded version of this model corresponds to the higher dimensional
generalization of the Randall-Sundrum 1-brane model with the brane location
at the orbifold fixed point $y=0$. The corresponding partial zeta function
is given by Eq. (\ref{FLs}). Now in this formula $\mathrm{J=L,R}$ for the
left and right sides of the brane, respectively, and $F=K$ for $\mathrm{J=R}$
and $F=I$ for $\mathrm{J=L}$. In addition, the replacement $%
n^{(j)}\rightarrow n^{\mathrm{(J)}}$ should be done with $n^{\mathrm{(R)}}=1$
and $n^{\mathrm{(L)}}=-1$. The integral representation (\ref{FLs}) for a
single brane partial zeta function is valid in the strip $-D_{1}-1<\mathrm{Re%
}\,s<-D_{1}$ and under the assumption that the function $\bar{F}_{\nu
}^{(j)}(u)$ has no real zeros. For the analytic continuation to $s=-1$ we
employ the asymptotic expansions of the modified Bessel functions for large
values of the argument \cite{abramowiz}. For $B_{j}\neq 0$ from these
expansions one has
\begin{equation}
\frac{F_{\nu }(u)}{\bar{F}_{\nu }^{(j)}(u)}\sim \frac{1}{B_{j}}%
\sum_{l=1}^{\infty }\frac{v_{l}^{(F,j)}}{u^{l}},  \label{Inuratas}
\end{equation}%
where the coefficients $v_{l}^{(F,j)}(\nu )$ are combinations of the
corresponding coefficients in the expansions for the functions $F_{\nu }(u)$
and $F_{\nu }^{\prime }(u)$. Note that one has the relation $%
v_{l}^{(K,j)}=(-1)^{l}v_{l}^{(I,j)}$, assuming that the coefficients in the
boundary conditions are the same for both sides of the brane. For the
nonzero modes along the internal space $\Sigma $\ we subtract and add to the
integrand in (\ref{FLs}) the $N$ leading terms of the corresponding
asymptotic expansion and exactly integrate the asymptotic part. For the zero
mode we first separate the integral over the interval $(0,1)$ and apply the
described procedure to the integral over $(1,\infty )$. As a result, the
corresponding function
\begin{equation}
\Phi _{j}^{\mathrm{(J)}}(s)=\frac{z_{j}^{D}B_{j}}{(4\pi )^{(D_{1}-1)/2}}%
\sum_{\beta }|\psi _{\beta }(X)|^{2}\zeta _{j\beta }^{\mathrm{(J)}}(s),
\label{FLsa}
\end{equation}%
is written in the form

\begin{eqnarray}
\Phi _{j}^{\text{\textrm{(J)}}}(s) &=&-\frac{(4\pi )^{(1-D_{1})/2}n^{\text{%
\textrm{(J)}}}z_{j}^{D_{2}}}{\Gamma \left( -\frac{s}{2}\right) \Gamma
(\alpha _{s}+1)(\mu z_{j})^{s+1}}\Bigg\{\sum_{\beta }|\psi _{\beta }(X)|^{2}%
\Bigg[\delta _{0\lambda _{\beta }}B_{j}\int_{0}^{1}du\,u^{D_{1}+s}\frac{%
F_{\nu }(u)}{\bar{F}_{\nu }^{(j)}(u)}  \notag \\
&&+\int_{u_{\beta }}^{\infty }du\,u(u^{2}-\lambda _{\beta
}^{2}z_{j}^{2})^{\alpha _{s}}\left( B_{j}\frac{F_{\nu }(u)}{\bar{F}_{\nu
}^{(j)}(u)}-\sum_{l=1}^{N}\frac{v_{l}^{F,j}}{u^{l}}\right) \Bigg]%
-\sum_{l=1}^{N}\frac{|\psi _{0}(X)|^{2}v_{l}^{(F,j)}}{D_{1}+s-l+1}  \notag \\
&&+\frac{1}{2}\Gamma (\alpha _{s}+1)\sum_{l=1}^{N}\frac{%
v_{l}^{(F,j)}z_{j}{}^{D_{1}+s-l+1}}{\Gamma (l/2)}\Gamma \left( \frac{l}{2}%
-\alpha _{s}-1\right) \zeta _{\Sigma }\left( \frac{l}{2}-\alpha
_{s}-1,X\right) \Bigg\},  \label{FLs1}
\end{eqnarray}%
where $u_{\beta }=\lambda _{\beta }z_{j}+\delta _{0\lambda _{\beta
}}$. In Eq.~(\ref{FLs1}) we have introduced the local spectral
zeta function associated with the massless laplacian defined on
the internal subspace $\Sigma $:
\begin{equation}
\zeta _{\Sigma }(z,X)=\sideset{}{'}{\sum}_{\beta }|\psi _{\beta
}(X)|^{2}\lambda _{\beta }^{-2z},  \label{localzeta}
\end{equation}%
where the prime on the summation sign means that the zero mode should be
omitted. Both integrals in Eq. (\ref{FLs1}) are finite at $s=-1$ for $%
N\geqslant D_{1}-1$. For large values $\lambda _{\beta }$ the second
integral behaves as $\lambda _{\beta }^{D_{1}+s-N}$ and the series over $%
\beta $ in Eq. (\ref{FLs1}) is convergent at $s=-1$ for $N>D-1$. For these
values $N$ the poles at $s=-1$ are contained only in the last two terms on
the right.

The zero mode part has a simple pole at $s=-1$ presented by the summand $%
l=D-1$ of the second sum in figure braces. The pole part corresponding to
the nonzero modes is extracted from the pole structure of the local zeta
function (\ref{localzeta}). The latter is given by the formula%
\begin{equation}
\Gamma (z)\zeta _{\Sigma }(z,X)|_{z=p}=\frac{C_{D_{2}/2-p}(X)}{z-p}+\Omega
_{p}(X)+\mathcal{\cdots },  \label{zetapoles}
\end{equation}%
where $p$ is a half integer, the coefficients $C_{D_{2}/2-p}(X)$ are related
to the Seeley-DeWitt or heat kernel coefficients for the corresponding
non-minimal laplacian, and the dots denote the terms vanishing at $z=p$. In
the way similar to that used in Ref. \cite{Flac03}, it can be seen that the
coefficients $C_{p}(X)$ are related to the corresponding coefficients $%
C_{p}(X,m)$ for the massive zeta function%
\begin{equation}
\zeta _{\Sigma }(s,X;m)=\sum_{\beta }\frac{|\psi _{\beta }(X)|^{2}}{(\lambda
_{\beta }^{2}+m^{2})^{s}},  \label{zetSigsXm}
\end{equation}%
by the formula%
\begin{equation}
C_{p}(X)=C_{p}(X,0)-|\psi _{0}(X)|^{2}\delta _{p,D_{2}/2}.  \label{CpX}
\end{equation}%
The VEV of the energy density on a single brane is derived from%
\begin{equation}
\varepsilon _{j}^{\mathrm{(J)}}=-n^{\mathrm{(J)}}k_{D}^{D}C_{j}\Phi _{j}^{%
\mathrm{(J)}}(s)|_{s=-1}.  \label{epsjJ}
\end{equation}%
By using relation (\ref{zetapoles}), the energy density is written as a sum
of pole and finite parts: $\varepsilon _{j}^{\mathrm{(J)}}=\varepsilon _{j,%
\mathrm{p}}^{\mathrm{(J)}}+\varepsilon _{j,\mathrm{f}}^{\mathrm{(J)}}$.
Laurent-expanding the expression on the right of Eq.~(\ref{FLs1}) near $s=-1$%
, one finds%
\begin{equation}
\varepsilon _{j,\text{\textrm{p}}}^{\text{\textrm{(J)}}}=-\frac{%
2k_{D}^{D}C_{j}}{(4\pi )^{D_{1}/2}(s+1)}\sum_{l=1}^{D}\frac{%
v_{l}^{(F,j)}z_{j}{}^{D-l}}{\Gamma (l/2)}\left[ C_{(D-l)/2}(X)+|\psi
_{0}(X)|^{2}\delta _{lD_{1}}\right]   \label{phi2Lpf1}
\end{equation}%
for the pole part, and

\begin{eqnarray}
\varepsilon _{j,\text{\textrm{f}}}^{\text{\textrm{(J)}}}
&=&2k_{D}^{D}z_{j}^{D_{2}}C_{j}\beta _{D_{1}}\sum_{\beta }|\psi _{\beta
}(X)|^{2}\left[ \delta _{0\lambda _{\beta }}B_{j}\int_{0}^{1}du\,u^{D-1}%
\frac{F_{\nu }(u)}{\bar{F}_{\nu }^{(j)}(u)}\right.   \notag \\
&&\left. +\int_{u_{\beta }}^{\infty }du\,u(u^{2}-\lambda _{\beta
}^{2}z_{j}^{2})^{D_{1}/2-1}\left( B_{j}\frac{F_{\nu }(u)}{\bar{F}_{\nu
}^{(j)}(u)}-\sum_{l=1}^{N}\frac{v_{l}^{(F,j)}}{u^{l}}\right) \right]   \notag
\\
&&-k_{D}^{D}z_{j}^{D_{2}}C_{j}\beta _{D_{1}}|\psi _{0}(X)|^{2}\Bigg\{%
\sideset{}{'}{\sum}_{l=1}^{N}\frac{2v_{l}^{(F,j)}}{D_{1}-l}-v_{D_{1}}^{(F,j)}%
\left[ 2\ln (\mu z_{j})+\psi \left( \frac{D_{1}}{2}\right) -\psi \left(
\frac{1}{2}\right) \right] \Bigg\}  \notag \\
&&+\frac{k_{D}^{D}C_{j}}{(4\pi )^{D_{1}/2}}\sum_{l=1}^{N}\frac{v_{l}^{(F,j)}%
}{\Gamma (l/2)}z_{j}^{D-l}\left\{ C_{(D-l)/2}(X)\left[ 2\ln \mu -\psi \left(
\frac{1}{2}\right) \right] +\Omega _{(l-D_{1})/2}(X)\right\}
\label{phi2Lpf2}
\end{eqnarray}%
for the finite part, with $\psi (x)$ being the diagamma function. In this
formula the prime on the summation sign means that the term with $l=D_{1}$
should be omitted and it is understood that $C_{p}(X)=0$ for $p<0$. In the
pole part the second term in the square brackets comes from the zero mode
along $\Sigma $ and this term is cancelled by the delta term on the right of
Eq. (\ref{CpX}).

The renormalization of the surface energy density can be done modifying the
procedure used previously for the renormalization of the Casimir energy in
the Randall-Sundrum model \cite{Flac01b,Gold00,Garr01} and in its
higher-dimensional generalizations with compact internal spaces \cite%
{Flac03,Flac03b}. The form of the counterterms needed for the
renormalization is determined by the pole part of the surface energy density
given by Eq. (\ref{phi2Lpf1}). For an internal manifold with no boundaries,
this part has the structure $\sum_{l=0}^{[(D-1)/2]}a_{l}^{\mathrm{(s)}%
}(z_{j}/L)^{2l}$. By taking into account that the intrinsic scalar curvature
$R_{j}$ for the brane at $y=j$ contains the factor $(z_{j}/L)^{2}$, we see
that the pole part can be absorbed by adding to the brane action
counterterms of the form%
\begin{equation}
\int d^{D}x\sqrt{|h|}\sum_{l=0}^{[(D-1)/2]}b_{l}^{\mathrm{(s)}}R_{j}^{l},
\label{Counterterms}
\end{equation}%
where the square brackets in the upper limit of summation mean the integer
part of the enclosed expression. By taking into account that there is the
freedom to perform finite renormalizations, we see that the renormalized
surface energy density on a single brane is given by the formula%
\begin{equation}
\varepsilon _{j}^{\mathrm{(J,ren)}}=\varepsilon _{j,\mathrm{f}}^{\mathrm{(J)}%
}+\sum_{l=0}^{[(D-1)/2]}c_{l}^{\mathrm{(s)}}(z_{j}/L)^{2l}.  \label{epsjren}
\end{equation}%
The coefficients $c_{l}^{\mathrm{(s)}}$ in the finite renormalization terms
are not computable within the framework of the model under consideration and
their values should be fixed by additional renormalization conditions.

The total surface energy density for a single brane at $y=j$ is obtained by
summing the contributions from the left and right sides: $\varepsilon _{j}^{{%
\mathrm{(LR)}}}=\varepsilon _{j}^{\mathrm{(L)}}+\varepsilon _{j}^{\mathrm{(R)%
}}$. In formulas (\ref{phi2Lpf1}), (\ref{phi2Lpf2}) we should take $F=I$ for
$\mathrm{J=L}$ and $F=K$ for $\mathrm{J=R}$. Now we see that, assuming the
same boundary conditions on both sides of the brane, the coefficients $%
C_{p}(X,0)$ enter into the sum of pole terms in the form%
\begin{equation}
2\sum_{l=1}^{[D/2]}\frac{v_{2l}^{(I,j)}}{\Gamma (l)}%
z_{j}{}^{D-2l}C_{D/2-l}(X,0).  \label{Cinsumpole}
\end{equation}%
If the internal manifold contains no boundaries and $D$ is an odd number,
one has $C_{D/2-l}(X,0)=0$ and, hence, the pole parts coming from the left
and right sides cancel out. For a one parameter internal space of size $L$
the surface energy density on the brane at $y=j$ is a function on the ratio $%
L/z_{j}$ only. Note that in the case of the AdS bulk the corresponding
quantity does not depend on the brane position. To discuss the physics from
the point of view of an observer residing on the brane, it is convenient to
introduce rescaled coordinates%
\begin{equation}
x^{\prime M}=e^{-k_{D}j}x^{M},\;M=0,1,\ldots ,D-1.  \label{scaledCoord}
\end{equation}%
With this coordinates the warp factor \ in the metric is equal to 1 on the
brane and they are physical coordinates for an observer on the brane. For
this observer the physical size of the subspace $\Sigma $ is $%
L_{j}=Le^{-k_{D}j}$ and the corresponding KK masses are rescaled by the warp
factor: $\lambda _{\beta }^{(j)}=\lambda _{\beta }e^{k_{D}j}$. Now we see
that the surface energy density is a function on the ratio $L_{j}/(1/k_{D})$
of the physical size for the internal space (for an observer residing on the
brane) to the AdS curvature radius.

As an application of the general results presented above, we can consider a
simple example with $\Sigma =S^{1}$. In this case the bulk corresponds to
the $\mathrm{AdS}_{D+1}$ spacetime with one compactified dimension $X$. The
corresponding normalized eigenfunctions are $\psi _{\beta }(X)=e^{2\pi
i\beta X/L}/\sqrt{L}$ with $\beta =0,\pm 1,\pm 2,\ldots $, where $L$ is the
length of the compactified dimension. The surface energy density induced on
the brane is obtained from general formulas by the replacements
\begin{equation}
\sum_{\beta }|\psi _{\beta }(X)|^{2}\rightarrow \frac{2}{L}%
\sideset{}{'}{\sum}_{\beta =0}^{\infty },\quad \lambda _{\beta }\rightarrow
\frac{2\pi }{L}|\beta |,\quad D_{2}=1,  \label{replaceforS1}
\end{equation}%
where the prime means that the summand $\beta =0$ should be taken with the
weight 1/2. For the local zeta function from Eq. (\ref{localzeta}) one has $%
\zeta _{\Sigma }(s,X)=2L^{2s-1}\zeta _{R}(2s)/(2\pi )^{2s}$, where $\zeta
_{R}(z)$ is the Riemann zeta function. Now the only poles of the function $%
\Gamma (z)\zeta _{\Sigma }(z,X)$ are the points $z=0,1/2$. By using the
standard formulas for the gamma function and the Riemann zeta function (see,
for instance, \cite{abramowiz}), it can be seen that one has $C_{1/2}(X)=-1/L
$, $C_{0}(X)=1/2\sqrt{\pi }$, for the residues appearing in (\ref{zetapoles}%
) and
\begin{equation}  \label{OmegaS1}
\Omega _{0}(X)=\frac{\gamma -2\ln L}{L},\;\Omega _{\frac{1}{2}}(X)=\frac{%
\gamma +2\ln (L/4\pi )}{2\sqrt{\pi }}, \; \Omega
_{p}(X)=\frac{2L^{2p-1}}{(2\pi )^{2p}}\Gamma (p)\zeta
_{R}(2p),\;p\neq 0,\;\frac{1}{2},
\end{equation}%
for the finite parts, with $\gamma $ being the Euler constant.

\section{Two-brane geometry and induced cosmological constant}

\label{sec:2brane}

As it has been shown in Section \ref{sec:zetafunc}, the partial
zeta function related to the surface energy density on the brane
at $y=j$ is presented in the form (\ref{zet12}), where the second
term on the right is finite at the physical point $s=-1$. For
two-brane geometry the VEV of the surface energy density on the
brane at $y=j$ is presented as the sum
\begin{equation}
\varepsilon _{j}^{{\mathrm{(s)}}}=\varepsilon _{j}^{{\mathrm{(J)}}}+\Delta
\varepsilon _{j}^{{\mathrm{(s)}}}.  \label{emt2pl2}
\end{equation}%
The first term on the right is the energy density induced on a single brane
when the second brane is absent. The second term is induced by the presence
of the second brane and is given by the formula
\begin{equation}
\Delta \varepsilon _{j}^{{\mathrm{(s)}}%
}=2C_{j}n^{(j)}(k_{D}z_{j})^{D}B_{j}^{2}\beta _{D_{1}}\sum_{\beta }|\psi
_{\beta }(X)|^{2}\int_{\lambda _{\beta }}^{\infty }du\,u(u^{2}-\lambda
_{\beta }^{2})^{D_{1}/2-1}\Omega _{j\nu }(uz_{a},uz_{b}).  \label{emt2pl3}
\end{equation}%
As we consider the region $a\leqslant y\leqslant b$, the energy
desnity (\ref{emt2pl3}) is located on the surface $y=a+0$ for the
left brane
and on the surface $y=b-0$ for the right brane. Consequently, in formula (%
\ref{emt2pl2}) we take $\mathrm{J=R}$ for $j=a$ and $\mathrm{J=L}$ for $j=b$%
. The energy densities on the surfaces $y=a-0$ and $y=b+0$ are the same as
for the corresponding single brane geometry. The expression on the right of
Eq. (\ref{emt2pl3}) is finite for all nonzero distances between the branes
and is not touched by the renormalization procedure. For a given value of
the AdS energy scale $k_{D}$ and one parameter manifold $\Sigma $ with the
length scale $L$, it is a function on the ratios $z_{b}/z_{a}$ and $L/z_{a}$%
. The first ratio is related to the proper distance between the branes, $%
z_{b}/z_{a}=\exp [k_{D}(b-a)]$, and the second one is the ratio of the size
of the internal space, measured by an observer residing on the brane at $y=a$%
, to the AdS curvature radius $k_{D}^{-1}$.

For the comparison with the case of the bulk spacetime $\mathrm{AdS}%
_{D_{1}+1}$ when the internal space is absent, it is useful in addition to
the VEV (\ref{emt2pl3}) to consider the corresponding quantity integrated
over the subspace~$\Sigma $:
\begin{equation}
\Delta \varepsilon _{D_{1}j}^{{\mathrm{(s)}}}=\int_{\Sigma }d^{D_{2}}X\sqrt{%
\gamma }\,\Delta \varepsilon _{j}^{{\mathrm{(s)}}%
}e^{-D_{2}k_{D}j}=e^{-D_{2}k_{D}j}\sum_{\beta }\Delta \varepsilon _{j\beta
}^{{\mathrm{(s)}}},  \label{phi2integrated}
\end{equation}%
where $\Delta \varepsilon _{j\beta }^{{\mathrm{(s)}}}$ is defined by the
relation
\begin{equation}
\Delta \varepsilon _{j}^{{\mathrm{(s)}}}=\sum_{\beta }|\psi _{\beta
}(X)|^{2}\Delta \varepsilon _{j\beta }^{{\mathrm{(s)}}}.  \label{DeltDef}
\end{equation}%
Comparing this integrated VEV with the corresponding formula from Ref. \cite%
{Saha04surf}, we see that the contribution of the zero KK mode ($\lambda
_{\beta }=0$) in Eq. (\ref{phi2integrated}) differs from the VEV of the
energy density in the bulk $\mathrm{AdS}_{D_{1}+1}$ by the order of the
modified Bessel functions: for the latter case $\nu \rightarrow \nu _{1}$
with $\nu _{1}$ defined by Eq. (\ref{nu}) with the replacement $D\rightarrow
D_{1}$.

Now we turn to the investigation of the part (\ref{emt2pl3}) in the surface
energy density in asymptotic regions of the parameters. For large values of
AdS radius compared with the interbrane distance, $k_{D}(b-a)\ll 1$, the
main contribution to the integral on the right of Eq. (\ref{emt2pl3}) comes
from large values of $uz_{a}\sim \lbrack k_{D}(b-a)]^{-1}$. Assuming that $%
\tilde{B}_{a}/(b-a)$ and $m(b-a)$ are fixed, we see that the order of the
modified Bessel functions is large. Replacing these functions by their
uniform asymptotic expansions for large values of the order \cite{abramowiz}%
, it can be seen that to the leading order the corresponding surface energy
on the branes in the bulk geometry $R^{(D_{1}-1,1)}\times \Sigma $ is
obtained.

For large KK masses along $\Sigma $, $z_{a}\lambda _{\beta }\gg 1$, $\lambda
_{\beta }\gg 1$, we can replace the modified Bessel functions by the
corresponding asymptotic expansions for large values of the argument. For
the contribution of a given KK mode to the leading order this gives%
\begin{equation}
\Delta \varepsilon _{j\beta }^{{\mathrm{(s)}}} \approx
4n^{(j)}k_{D}^{D}z_{j}^{D+1}B_{j}^{2}C_{j}\beta _{D_{1}}\int_{\lambda
_{\beta }}^{\infty }du\,u^{2} \frac{%
(A_{j}^{2}-B_{j}^{2}u^{2}z_{j}^{2})^{-1}(u^{2}-\lambda _{\beta }^{2})^{\frac{%
D_{1}}{2}-1}}{c_{a}(uz_{a})c_{b}(uz_{b})e^{2u(z_{b}-z_{a})}-1},
\label{phi22pllargelamb}
\end{equation}%
where
\begin{equation}
c_{j}(u)=\frac{A_{j}-n^{(j)}B_{j}u}{A_{j}+n^{(j)}B_{j}u},\quad j=a,b.
\label{cj1}
\end{equation}%
If in addition one has the condition $\lambda _{\beta }(z_{b}-z_{a})\gg 1$,
the dominant contribution into the $u$-integral comes from the lower limit
and we have the formula%
\begin{equation}
\Delta \varepsilon _{j\beta }^{{\mathrm{(s)}}} \approx \frac{%
2n^{(j)}B_{j}^{2}C_{j}}{A_{j}^{2}-(\lambda _{\beta }z_{j}B_{j})^{2}}\frac{%
k_{D}^{D}z_{j}^{D+1}}{(4\pi )^{D_{1}/2}} \frac{\lambda _{\beta
}^{D_{1}/2+1}e^{-2\lambda _{\beta }(z_{b}-z_{a})}}{c_{a}(\lambda _{\beta
}z_{a})c_{b}(\lambda _{\beta }z_{b})(z_{b}-z_{a})^{D_{1}/2}}.
\label{epsjlargeKK1}
\end{equation}%
In particular, for sufficiently small length scale of the internal
space this formula is valid for all nonzero KK masses and the main
contribution to the surface densities comes from the zero KK mode.
In the opposite limit of large internal space, to the leading
order we obtain the corresponding result for parallel branes in
${\mathrm{AdS}}_{D+1}$ bulk \cite{Saha04surf}.

For small interbrane distances, $k_{D}(b-a)\ll 1$, which is equivalent to $%
z_{b}/z_{a}-1\ll 1$, the main contribution into the integral in Eq. (\ref%
{emt2pl3}) comes from large values $u$ and to the leading order we obtain
formula (\ref{phi22pllargelamb}). If in addition one has $\lambda _{\beta
}(z_{b}-z_{a})\ll 1$ or equivalently $\lambda _{\beta }^{(a)}(b-a)\ll 1$, we
can put in this formula $\lambda _{\beta }=0$. Assuming $(b-a)\ll |\tilde{B}%
_{j}/A_{j}|$, for $\tilde{B}_{j}\neq 0$ to the leading order one finds%
\begin{equation}
\Delta \varepsilon _{j\beta }^{{\mathrm{(s)}}}\approx -4k_{D}\sigma
_{j}n^{(j)}\frac{C_{j}\Gamma \left( \frac{D_{1}-1}{2}\right) \zeta
_{R}(D_{1}-1)}{(4\pi )^{(D_{1}+1)/2}(b-a)^{D_{1}-1}}e^{D_{2}k_{D}j},
\label{epsjsmalldist}
\end{equation}%
where $\sigma _{j}=1$ for $|B_{a}/A_{a}|,|B_{b}/A_{b}|\gg k_{D}(b-a)$, and $%
\sigma _{j}=2^{2-D}-1$ for $|B_{j}/A_{j}|\gg k_{D}(b-a)$ and $B_{l}/A_{l}=0$%
, with $l=b$ for $j=a$ and $l=a$ for $j=b$. We see that for small interbrane
distances the sign of the induced surface energy density is determined by
the coefficient $C_{j}$ and this sign is different for two cases of $\sigma
_{j}$.

Now we consider the limit $\lambda _{\beta }z_{b}\gg 1$ assuming that $%
\lambda _{\beta }z_{a}\lesssim 1$. Using the asymptotic formulas for the
Bessel modified functions containing in the argument $z_{b}$, for the
contribution of a given nonzero KK mode we find the following results%
\begin{eqnarray}
\Delta \varepsilon _{a\beta }^{{\mathrm{(s)}}} &\approx &\frac{%
k_{D}^{D}z_{a}^{D}B_{a}^{2}C_{a}}{2^{D_{1}}\pi ^{D_{1}/2-1}}\frac{(\lambda
_{\beta }/z_{b})^{D_{1}/2}e^{-2\lambda _{\beta }z_{b}}}{c_{b}(\lambda
_{\beta }z_{b})\bar{K}_{\nu }^{(a)2}(\lambda _{\beta }z_{a})},
\label{epslargezb1} \\
\Delta \varepsilon _{b\beta }^{{\mathrm{(s)}}} &\approx &-\frac{%
k_{D}^{D}z_{b}^{D+1}B_{b}^{2}C_{b}}{2^{D_{1}-1}\pi
^{D_{1}/2-1}z_{b}^{D_{1}/2}}\frac{\lambda _{\beta }^{D_{1}/2+1}e^{-2\lambda
_{\beta }z_{b}}}{(A_{b}+\lambda _{\beta }z_{b}B_{b})^{2}} \frac{\bar{I}_{\nu
}^{(a)}(\lambda _{\beta }z_{a})}{\bar{K}_{\nu }^{(a)}(\lambda _{\beta }z_{a})%
}.  \label{epslargezb2}
\end{eqnarray}%
This limit corresponds to the interbrane distances much larger compared with
the AdS curvature radius and with the inverse KK masses measured by an
observer on the left brane: $(b-a)\gg 1/k_{D},1/\lambda _{\beta }^{(a)}$.
For a single parameter manifold $\Sigma $ with length scale $L$ and $%
(b-a)\gg L_{a}$ these conditions are satisfied for all nonzero KK modes.

In the limit $z_{a}\lambda _{\beta }\ll 1$ for fixed $z_{b}\lambda _{\beta }$%
, by using the asymptotic formulas for the modified Bessel functions for
small values of the argument and assuming $|A_{a}|\neq |B_{a}|\nu $, one
finds%
\begin{eqnarray}
\Delta \varepsilon _{a\beta }^{{\mathrm{(s)}}} &\approx &\frac{%
k_{D}^{D}z_{a}^{D+2\nu }B_{a}^{2}C_{a}\beta _{D_{1}}}{2^{2\nu -3}\Gamma
^{2}(\nu )(A_{a}-\nu B_{a})^{2}}\int_{\lambda _{\beta }}^{\infty
}du\,u^{2\nu +1}(u^{2}-\lambda _{\beta }^{2})^{D_{1}/2-1}\frac{\bar{K}_{\nu
}^{(b)}(uz_{b})}{\bar{I}_{\nu }^{(b)}(uz_{b})},  \label{epssmallza1} \\
\Delta \varepsilon _{b\beta }^{{\mathrm{(s)}}} &\approx &-\frac{k_{D}^{D_{1}}%
}{c_{a}(\nu )}\left( \frac{z_{a}}{z_{b}}\right) ^{2\nu
}k_{D}^{D_{2}}z_{b}^{D_{2}}f_{\nu \beta }^{(b)},  \label{epssmallza2}
\end{eqnarray}%
where we have introduced the notation%
\begin{equation}
f_{\nu \beta }^{(b)}=\frac{4B_{b}^{2}C_{b}\beta _{D_{1}}}{2^{2\nu }\nu
\Gamma ^{2}(\nu )}\int_{\lambda _{\beta }z_{b}}^{\infty }du\,\frac{u^{2\nu
+1}}{\bar{I}_{\nu }^{(b)2}(u)}(u^{2}-\lambda _{\beta
}^{2}z_{b}^{2})^{D_{1}/2-1}.  \label{fnubet}
\end{equation}%
For $|A_{a}|=|B_{a}|\nu $ we should take into account the next terms in the
corresponding expansions of the modified Bessel functions. The integral in
Eq. (\ref{epssmallza1}) is negative for small values of the ratio $%
A_{b}/B_{b}$ and is positive for large values of this ratio. As it follows
from Eq. (\ref{epssmallza2}), for large interbtane separations the sign of
the quantity $\Delta \varepsilon _{b}^{{\mathrm{(s)}}}$ is determined by the
combination $(B_{a}^{2}\nu ^{2}-A_{a}^{2})C_{b}$ of the coefficients in the
boundary conditions. In the limit under consideration the KK masses measured
by an observer on the brane at $y=a$ are much less than the AdS energy
scale, $\lambda _{\beta }^{(a)}\ll k_{D}$, and the interbrane distance is
much larger than the AdS curvature radius. In particular, substituting $%
\lambda _{\beta }=0$, from these formulas we obtain the asymptotic behavior
for the contribution of the zero mode to the surface energy density induced
by the second brane in the limit $z_{a}/z_{b}\ll 1$. Now combining the
corresponding result with formulas (\ref{epslargezb1}), (\ref{epslargezb2}),
we see that under the conditions $(b-a)\gg 1/k_{D},L_{a}$, and $%
L_{a}k_{D}\gtrsim 1$, the contribution of the nonzero KK modes along $\Sigma
$ is suppressed with respect to the contribution of the zero mode by the
factor $(z_{b}/z_{a})^{2\nu +D_{1}/2+\delta _{j}^{b}}\exp (-2\lambda _{\beta
}z_{b})$.

From the analysis given above it follows that in the limit when the right
brane tends to the AdS horizon, $z_{b}\rightarrow \infty $, the energy
density $\Delta \varepsilon _{a\beta }^{{\mathrm{(s)}}}$ vanishes as $%
e^{-2\lambda _{\beta }z_{b}}/z_{b}^{D_{1}/2}$ for the nonzero KK mode along $%
\Sigma $ and as $z_{b}^{-D_{1}-2\nu }$ for the zero mode. The energy density
on the right brane, $\Delta \varepsilon _{b\beta }^{{\mathrm{(s)}}}$,
vanishes as $z_{b}^{D_{2}+D_{1}/2+1}e^{-2\lambda _{\beta }z_{b}}$ for the
nonzero KK mode and behaves like $z_{b}^{D_{2}-2\nu }$ for the zero mode. In
the limit when the left brane tends to the AdS boundary, $z_{a}\rightarrow 0$%
, the contribution of a given KK mode vanishes as $z_{a}^{D+2\nu }$ for $%
\Delta \varepsilon _{a\beta }^{{\mathrm{(s)}}}$ and as $z_{a}^{2\nu }$ for $%
\Delta \varepsilon _{b\beta }^{{\mathrm{(s)}}}$. For small values of the AdS
curvature radius corresponding to strong gravitational fields, assuming $%
\lambda _{\beta }z_{a}\gg 1$ and $\lambda _{\beta }(z_{b}-z_{a})\gg 1$, we
can estimate the contribution of the nonzero KK modes to the induced energy
densities by formula (\ref{epsjlargeKK1}). In particular, for the case of a
single parameter internal space with the length scale $L$, under the assumed
conditions the length scale of the internal space measured by an observer on
the brane at $y=a$ is much smaller compared to the AdS curvature radius, $%
L_{a}\ll k_{D}^{-1}$. If $L_{a}\gtrsim k_{D}^{-1}$ one has
$\lambda _{\beta }z_{a}\lesssim 1$ and to estimate the
contribution of the induced surface densities we can use formulas
(\ref{epslargezb1}) and (\ref{epslargezb2}), and the suppression
is stronger compared with the previous case. For the zero KK mode,
under the condition $k_{D}(b-a)\gg 1$ we have $z_{a}/z_{b}\ll 1 $
and to the leading order the corresponding energy densities are
described by relations (\ref{epssmallza1}) and
(\ref{epssmallza2}). From these formulae it follows that the
induced energy densities integrated over the internal space behave
as $k_{D}^{D_{1}+1}\exp [(D_{1}\delta _{j}^{a}+2\nu )k_{D}(a-b)]$
for the brane at $y=j$ and are exponentially suppressed.

Introducing the rescaled coordinates defined by Eq. (\ref{scaledCoord}),
after the Kaluza-Klein reduction of the higher dimensional Hilbert action,
by the way similar to that in the Randall-Sundrum braneworld, it can be seen
that effective $D_{1}$-dimensional Newton's constant $G_{D_{1}j}$ measured
by an observer on the brane at $y=j$ and the fundamental $(D+1)$-dimensional
Newton's constant $G_{D+1}$ are related by the formula
\begin{equation}
G_{D_{1}j}=\frac{(D-2)k_{D}G_{D+1}}{V_{\Sigma j}\left[ e^{(D-2)k_{D}(b-a)}-1%
\right] }e^{(D-2)k_{D}(b-j)},\;V_{\Sigma j}=e^{-D_{2}k_{D}j}\int d^{D_{2}}X%
\sqrt{\gamma },  \label{GDj}
\end{equation}%
where $V_{\Sigma j}$ is the volume of the internal space measured by the
same observer. In the orbifolded version of the model an additional factor 2
appears in the denominator of the expression on the right. Formula (\ref{GDj}%
) explicitly shows two possibilities for the hierarchy generation by the
redshift and large volume effects. Note that the ratio of the Newton
constants on the branes, $G_{D_{1}b}/G_{D_{1}a}=e^{(2-D_{1})k_{D}(b-a)}$, is
the same as in the model without an internal space. For large interbrane
distances one has $G_{D_{1}a}\sim k_{D}G_{D+1}/V_{\Sigma a}$, $%
G_{D_{1}b}\sim k_{D}G_{D+1}e^{(2-D)k_{D}(b-a)}/V_{\Sigma b}$, and the
gravitational interactions on the brane $y=b$ are exponentially suppressed.
This feature is used in the Randall-Sundrum model to address the hierarchy
problem. As we will see below this mechanism also allows to obtain a
naturally small cosmological constant generated by the vacuum quantum
fluctuations (for the discussion of the cosmological constant problem within
the framework of braneworld models see references given in \cite{Saha04surf}%
).

As we have already mentioned, surface energy density (\ref{emt2pl3})
corresponds to the gravitational source of the cosmological constant type
induced on the brane at $y=j$ by the presence of the second brane. For an
observer living on the brane at $y=j$ the corresponding effective $D_{1}$%
-dimensional cosmological constant is determined by the relation
\begin{equation}
\Lambda _{D_{1}j}=8\pi G_{D_{1}j}\Delta \varepsilon _{D_{1}j}^{{\mathrm{(s)}}%
}=8\pi M_{D_{1}j}^{2-D_{1}}\Delta \varepsilon
_{D_{1}j}^{{\mathrm{(s)}}},  \label{effCC}
\end{equation}%
where $M_{D_{1}j}$ is the $D_{1}$-dimensional effective Planck mass scale
for the same observer and $\Delta \varepsilon _{D_{1}j}^{{\mathrm{(s)}}}$ is
defined by Eq. (\ref{phi2integrated}). Denoting by $M_{D+1}$ the fundamental
$(D+1)$-dimensional Planck mass, $G_{D+1}=M_{D+1}^{1-D}$, from Eq. (\ref{GDj}%
) one has the following relation%
\begin{equation}
\left( \frac{M_{D_{1}j}}{M_{D+1}}\right) ^{D_{1}-2}=\frac{%
(z_{b}/z_{a})^{D-2}-1}{(D-2)(z_{b}/z_{j})^{D-2}}\frac{V_{\Sigma j}}{k_{D}}%
M_{D+1}^{D_{2}+1},  \label{Planckhierarchy}
\end{equation}%
for the ratio of the effective and fundamental Planck scales. By using the
asypmtotic relations given above, for large interbrane distances one obtains
the following estimate for the ratio of the induced cosmological constant (%
\ref{effCC}) to the corresponding Planck scale quantity in the brane
universe:%
\begin{equation}
h_{j}\equiv \frac{\Lambda _{D_{1}j}}{8\pi G_{D_{1}j}M_{D_{1}j}^{D_{1}}}\sim
\left( \frac{k_{D}^{D_{1}-1}}{V_{\Sigma j}M_{D+1}^{D-1}}\right) ^{\frac{D_{1}%
}{D_{1}-2}}\exp \Big[k_{D}(a-b)\Big(2\nu +D_{1}+\frac{D_{2}D_{1}}{D_{1}-2}%
\delta _{j}^{b}\Big)\Big].  \label{LambDjest1}
\end{equation}%
For the model without an internal space this ratio is of the same order of
magnitude for both branes.

In the higher dimensional version of the Randall-Sundrum braneworld the
brane at $z=z_{b}$ corresponds to the visible brane. For large interbrane
distances, by taking into account Eq. (\ref{epssmallza2}), for the ratio of
the induced cosmological constant (\ref{effCC}) to the Planck scale quantity
in the corresponding brane universe one obtains%
\begin{equation}
h_{b}\approx -\frac{1}{c_{a}(\nu )}\left( \frac{k_{D}}{M_{D_{1}b}}\right)
^{D_{1}}\left( \frac{z_{a}}{z_{b}}\right) ^{2\nu }\sum_{\beta }f_{\nu \beta
}^{(b)},  \label{LambDb}
\end{equation}%
where the function $f_{\nu \beta }^{(b)}$ is defined by Eq. (\ref{fnubet}).

Using relation (\ref{Planckhierarchy}) with $j=b$, we can express the
corresponding interbrane distance in terms of the ratio of the Planck scales
\begin{equation}
\frac{z_{a}}{z_{b}}\approx \left[ M_{D+1}^{D_{2}+1}\frac{V_{\Sigma b}}{k_{D}}%
\left( \frac{M_{D+1}}{M_{D_{1}b}}\right) ^{D_{1}-2}\right] ^{\frac{1}{D-2}}.
\label{Dist}
\end{equation}%
Substituting this into Eq. (\ref{LambDb}), for the ratio of the cosmological
constant on the brane at $j=b$ to the corresponding Planck scale quantity
one finds%
\begin{equation}
h_{b}\approx -\frac{1}{c_{a}(\nu )}\left( \frac{k_{D}}{M_{D+1}}\right)
^{D_{1}-\tilde{\nu}}\left( V_{\Sigma b}M_{D+1}^{D_{2}}\right) ^{\tilde{\nu}%
}\left( \frac{M_{D+1}}{M_{D_{1}b}}\right) ^{D_{1}+\tilde{\nu}%
(D_{1}-2)}\sum_{\beta }f_{\nu \beta }^{(b)},  \label{Lambonb}
\end{equation}%
with $\tilde{\nu}=2\nu /(D-2)$. The higher dimensional Planck mass $M_{D+1}$
and AdS inverse radius $k_{D}$ are two fundamental energy scales in the
theory which in the Randall-Sundrum model are usually assumed to be of the
same order, $k_{D}\sim M_{D+1}$. In this case one obtains the induced
cosmological constant which is exponentially suppressed compared with the
corresponding Planck scale quantity on the visible brane. In the model with $%
D_{1}=4$, $k_{D}\sim M_{D+1}\sim 1$ TeV, $M_{D_{1}b}=M_{{\mathrm{Pl}}}\sim
10^{16}$ TeV, assuming that the compactification scale on the visible brane
is close to the fundamental Planck scale, $V_{\Sigma b}M_{D+1}^{D_{2}}\sim 1$%
, for the ratio of the induced cosmological constant to the Planck scale
quantity on the visible brane we find the estimate $h_{b}\sim 10^{-32(2+%
\tilde{\nu})}$. From (\ref{Dist}) one has $k_{D}(b-a)\approx
74/(D_{2}+2)$ and the corresponding interbrane distances
generating the required hierarchy between the electroweak and
Planck scales are smaller than those
for the model without an internal space. In the model proposed in Ref. \cite%
{Flac03b}, a separation between the fundamental Planck scale and curvature
scale is assumed: $k_{D}\sim M_{D+1}z_{a}/z_{b}\sim 1$ TeV. Under the
assumption $V_{\Sigma b}M_{D+1}^{D_{2}}\sim 1$, in this model we have $%
h_{b}\sim 10^{-64[1+\nu /(D+1)]}$ and $k_{D}(b-a)\approx 74/(D_{2}+3)$.

\section{Conclusion}

\label{sec:Conc}

We have investigated the expectation value of the surface
energy-momentum tensor induced by the vacuum fluctuations of a
bulk scalar field with an arbitrary curvature coupling parameter
satisfying Robin boundary conditions on two parallel branes in
background spacetime $\mathrm{AdS}_{D_{1}+1}\times \Sigma $ with a
warped internal space $\Sigma $. Vacuum stresses on the brane are
the same for both subspaces and the energy-momentum tensor on the
brane corresponds to the source of the cosmological constant type
in the brane universe. It is remarkable that the latter property
is valid also for the more general model with the metric $g_{\mu
\sigma }$ instead of $\eta _{\mu \sigma }$ in line element
(\ref{metric}). As an regularization procedure for the surface
energy density we employ the zeta function technique. The
corresponding zeta function is presented as the sum of single
brane and second brane induced parts. The latter is finite at the
physical point and the further analytical continuation is
necessary for the first term only. As the first step we subtract
and add to the integrand the leading terms of the corresponding
asymptotic expansion for large values of the argument and
explicitly integrate the asymptotic part. Further, for the
regularization of the sum over the modes along the internal space
we use the local zeta function related to these modes. By making
use of the formula for the pole structure of this function, we
have presented the energy density on a single brane as the sum of
the pole and finite parts. The pole parts in the surface energy
density are absorbed by adding to the brane action the
counterterms having the structure given by Eq.
(\ref{Counterterms}). The renormalized energy density on the
corresponding surface of a single brane is determined by formula
(\ref{epsjren}), where the second term on the right presents the
finite renormalization part. The coefficients in this part cannot
be determined within the model under consideration and their
values should be fixed by additional renormalization conditions
which relate them to observables.

Unlike to the single brane part, the surface energy density induced by the
presence of the second brane contains no renormalization ambiguities and is
investigated in Section \ref{sec:2brane}. This part is given by formula (\ref%
{emt2pl3}). We have investigated the induced energy density in
various asymptotic regions for the parameters of the model. In the
model under discussion the hierarchy between the fundamental
Planck scale and the effective Planck scale in the brane universe
is generated by the combination of redshift and large volume
effects. The corresponding effective Newton's constant on the
brane at $y=j$ is related to the higher-dimensional fundamental
Newton's constant by formula (\ref{GDj})\ and for large interbrane
separations is exponentially small on the brane $y=b$. We show
that this mechanism also allows obtaining a naturally small
cosmological constant generated by the vacuum quantum fluctuations
of a bulk scalar. For large interbrane distances the ratio of the
induced cosmological constant to the the corresponding Planck
scale quantity in the brane universe is estimated by formula
(\ref{LambDjest1}) \ and is exponentially small. For the visible
brane in the higher dimensional generalization of the
Randall-Sundrum two-brane model, this ratio is given
in terms of the effective and fundamental Planck masses by Eq.~(\ref{Lambonb}%
). We have considered two classes of models with the compactification scale
on the visible brane close to the fundamental Planck scale. For the first
one the higher dimensional Planck mass and the AdS inverse radius are of the
same order and in the second one a separation between these scales is
assumed. In both cases the corresponding interbrane distances generating the
hierarchy between the electroweak and Planck scales are smaller than those
for the model without an internal space and the required suppression of the
cosmological constant is obtained without fine tuning.

\section*{Acknowledgments}

The work was supported by the Armenian Ministry of Education and Science
Grant No. 0124 and by PVE/CAPES Program.

\end{document}